\documentclass[twocolumn,superscriptaddress,aps,prl]{revtex4}
\usepackage{graphicx}

\begin{document}

\title{Shapes of hydrophobic thick membranes}

\author{Trinh X. Hoang}
\affiliation{Department of Physics, University of Maryland, College Park MD
20742, USA}
\affiliation{Institute of Physics,
Vietnam Academy of Science and Technology, 10 Dao Tan, Hanoi 10000, Vietnam}

\author{Jayanth R. Banavar}
\affiliation{Department of Physics, University of Maryland, College Park MD
20742, USA}

\author{Amos Maritan}
\affiliation{
Dipartimento di Fisica `G. Galilei', Universit\`a di Padova \&
CNISM, unit\`a di Padova \& INFN, sezione di Padova, Via Marzolo 8, 35131
Padova, Italy
}


\begin{abstract}
We introduce and study the behavior of a tethered membrane of {\it non-zero
thickness} embedded in three dimensions subject to an effective self-attraction
induced by hydrophobicity arising from the tendency to minimize the area
exposed to a solvent. The phase behavior and the nature of the folded
conformations are found to be quite distinct in the small and large solvent
size regimes. We demonstrate spontaneous symmetry-breaking with the membrane
folding along a preferential axis, when the solvent molecules are small
compared to the membrane thickness. For large solvent molecule size, a local
crinkling mechanism effectively shields the membrane from the solvent, even in
relatively flat conformations. We discuss the binding/unbinding transition of a
membrane to a wall that serves to shield the membrane from the solvent.
\end{abstract}

\maketitle

\section{Introduction}

The structure and dynamics of sheets or surfaces are important in diverse
contexts in everyday life, in cell biology, in field  theory and quantum
gravity, and in condensed matter physics \cite{Nelson,Lipowsky}.  Previous
studies have concentrated on idealized surfaces - the thickness is neglected
principally to facilitate an analytical approach (see the excellent work by
David, Duplantier, Guitter and Wiese, reviewed in Ref.\cite{Wiese}) or
partially captured by tethering together, in a two dimensional array, hard
spheres, whose diameter equals the surface thickness. In the latter case, both
the steric and interactions induced by the environment or solvent are modeled
as standard two-body potentials which become singular (Dirac delta function)
interactions in the continuum limit \cite{BanavarJSP}. In this limit,
self-interacting manifolds, in 1 dimension corresponding to a polymer chain and
in 2 dimensions corresponding to a membrane, the subject of our study, cannot
be correctly described by two body interactions \cite{BanavarJSP}.  The
difficulty arises because a pair-wise interaction can only depend on the
distance between the two parts of the manifold. One is unable to account for
whether these two parts are near each other just because they lie on adjacent
parts of the manifold or whether they are distant along the manifold and thence
truly interacting. One could cure this problem by introducing some microscopic
cut-off such that nearby parts of the manifold, at a distance smaller than this
cut-off along the manifold, are defined to be non-interacting. This procedure
is expected not to affect the long distance behavior but it affects many
important intermediate scale effects of interest.

Extensive studies \cite{Maritan,Marenduzzo,Banavar} have shown that there are
vastly different behaviors between a conventional polymer modeled as a chain of
non-overlapping spheres and a tube due distinct symmetries in the two cases.
The latter can be imagined as a chain of coins whose symmetry is cylindrical.
Self-avoidance of a flexible tube can be imposed by using a three-body
potential \cite{Gonzalez}. A tube subject to maximal compaction takes on the
geometry of an optimal helix \cite{Maritan}, whose local curvature is equal to
the tube thickness. Strikingly, this geometry is adopted by $\alpha$-helices in
proteins.  On the other hand, a fcc lattice is an optimal arrangement for a
chain of spheres subject to compaction, provided the tether constraints do not
conflict with this arrangement. Surprisingly, chains of overlapping spheres
\cite{Magee} behave in a manner akin to a tube. This is because the overlap
between adjacent spheres along the chain result in a loss of spherical symmetry
and endow an anisotropy to the chain much as a tube does \cite{Nanomachines}.

The principal theme of this paper is to introduce a model of a thick
hydrophobic membrane that encapsulates the correct symmetry associated with its
intrinsic two dimensional nature. The implementation of the non-zero thickness
of the membrane is carried out through a four-body potential \cite{BanavarJSP}.
Unlike the model of tethered spheres, the membrane thickness in our model
provides a length scale that is independent of the discretization of the
membrane surface, and thus is free of the singularity problem in the continuum
limit.  The interaction with the solvent molecules is captured through
Hadwiger's theorem \cite{Hadwiger} as explained in \cite{Dietrich}: the
effective self-interactions of the membrane, induced by a short-range
solvent-membrane potential, can have only four distinct contributions. Two of
these interactions are proportional to the integrals of the mean curvature and
Gaussian curvature over the entire surface. The other two interactions pertain
to an entropic contribution arising from the volume excluded to the solvent
molecules \cite{Kamien} and an energy contribution measured by the area of the
membrane that is exposed to the solvent molecules.  The membrane thickness
already takes into account the mean curvature effects because, at each position
on the surface, the curvature radii have to be larger than the thickness
itself. The integral of the Gaussian curvature is constant if the membrane
topology is kept fixed.  Furthermore, because the membranes we consider are
open, the volume excluded to the solvent molecules is partly taken into account
by the buried area or equivalently by its complement, the exposed area. Thus
our focus here is on understanding how the area of the membrane exposed to the
solvent molecules influences the nature of the conformations of a hydrophobic
membrane immersed in a solvent. 
We show that distinct behaviors are obtained for different ratios between
the membrane thickness and the solvent molecule size, both in
the absence and presence of an inert wall, which serves to shield one side
of the membrane from the solvent.
We find several familiar conformations that have not been observed in previous
studies of conventional models.

\section{Mathematical description of thick membrane}

The notion of thickness can be introduced for a continuum surface in 3-D.
The position vector of a point on the surface is given by
$\mathbf{r}(\vec{s})$, where $\vec{s}=(s^1,s^2)\in \mathcal{D}$ defines
the local (curvilinear) coordinate frame on the surface. The two tangent
vectors are
\begin{equation}\label{tang}
    \partial_i \mathbf{r}(\vec{s})\equiv
    \frac{\partial \mathbf{r}(\vec{s})}{\partial s^i},\ \
    i=1,2\ . 
\end{equation}
The square of the distance between two neighboring positions on the
surface, $\vec{s}$, $\vec{s} + \mathrm{d}\vec{s}$ is given by
\begin{equation}\label{dist}
    \mathrm{d}\mathbf{r}^2=g_{ij}\mathrm{d}s^i\mathrm{d}s^j\ , 
\end{equation}
where
\begin{equation}\label{metr}
g_{ij}\equiv \partial_i \mathbf{r}(\vec{s})\cdot \partial_j
\mathbf{r}(\vec{s})\ , 
\end{equation}
is the metric tensor. We will assume
that the tangent vectors are linearly independent at all points on
the surface which is equivalent to
\begin{equation}\label{g}
  \Big(\partial_1 \mathbf{r}(\vec{s})\times\partial_2
    \mathbf{r}(\vec{s})\Big)^2 = \det(g_{ij})\equiv g \neq 0\ . 
\end{equation}
The surface element at position $\vec{s}$ is given by
\begin{equation}\label{surf}
    \mathrm{d}\mathcal{S}(\vec{s})\equiv \mathrm{d}^2s\
    \sqrt{g(\vec{s})}\ . 
\end{equation}
We also assume that the local coordinate frame is chosen such that the
normal to the surface
\begin{equation}\label{n}
    \hat{\mathbf{n}}(\vec{s})=
    \frac{\partial_1 \mathbf{r}(\vec{s})\times\partial_2
    \mathbf{r}(\vec{s})}{|\partial_1 \mathbf{r}(\vec{s})\times\partial_2
    \mathbf{r}(\vec{s})|}\ , 
\end{equation}
is always on one side of the surface. If the thickness of the surface is
$2\Delta$, then the radius of the sphere tangent at $\mathbf{r}(\vec{s})$ and
going through another point $\mathbf{r}(\vec{t})$ on the surface
($\vec{t}$ is a vector in the curvilinear coordinates),
\begin{equation}\label{thick}
    \mathcal{R}=\frac{[\mathbf{r}(\vec{t})-\mathbf{r}(\vec{s})]^2}
    {2\hat{\mathbf{n}}(\vec{s})\cdot[\mathbf{r}(\vec{t})-\mathbf{r}(\vec{s})]}
    \ , 
\end{equation}
cannot be smaller than $\Delta$ for all $\vec{s}$ and $\vec{t}$ (fig. 1).

For a triangulated surface with fixed connectivity corresponding to the
middle surface of a membrane, assume that the vertices
are denoted as $i,j,k\ldots$ whose positions are given by ${\bf r}_i, {\bf
r}_j, {\bf r}_k \ldots$. A triangle of vertices $i$, $j$, and $k$ is denoted by
$(ijk)$.  For each triangle $(ijk)$, we assign an orientation such that for the
completely flat surface the normal vector
\begin{equation}
\hat{\mathbf{n}}_{ijk}=\frac{(\mathbf{r}_j-\mathbf{r}_i)\times
(\mathbf{r}_k-\mathbf{r}_j)}{|(\mathbf{r}_j-\mathbf{r}_i)\times
(\mathbf{r}_k-\mathbf{r}_j)|}
\label{eq:normal}
\end{equation}
always points in the same, say up, direction. The thickness is
defined in analogy with the continuum case.  The radius of the sphere tangent
to the triangle $(ijk)$ at its the center of mass $\mathbf{r}_{ijk}\equiv
(\mathbf{r}_i+\mathbf{r}_j+\mathbf{r}_k)/3$ and going through the
node $\mathbf{r}_m$,
\begin{equation}\label{discrthick}
    \mathcal{R}_{ijkm}=\frac{(\mathbf{r}_m-\mathbf{r}_{ijk})^2}
    {2|\hat{\mathbf{n}}_{ijk}\cdot(\mathbf{r}_m-\mathbf{r}_{ijk})|},
\end{equation}
is constrained to be greater than or equal to $\Delta\ \forall (ijk), m\neq i,j,k$.

\section{Exposed area of thick membrane}

Assume that the solvent molecule is a spherical paint brush of radius $R$
(fig. 1). The exposed area of a membrane is defined as the membrane's surface
area accessible by this spherical paint brush.
Each membrane has two sides, $+$ and $-$, defined by two triangulated surfaces,
$\mathcal{S}_{+}$ and $\mathcal{S}_{-}$, at a distance $\Delta$ from the middle
surface.
We assign a normal to each node $\mathbf{r}_i$ of the middle surface as the
{\it average} normal of all the triangles sharing the
$i$-th node: \begin{equation}\label{avenormal}
\hat{\mathbf{n}}_{i}=\frac{\sum_{jk}
\hat{\mathbf{n}}_{ijk}}{|\sum_{jk} \hat{\mathbf{n}}_{ijk}|} \ , \nonumber
\end{equation}
where 
the sum is over all oriented triangles sharing the $i$-th node.
The number of such triangles is six in the bulk and less than
six at the boundary. Each $\hat{\mathbf{n}}_{i}$, by definition, is
oriented in the direction of $\mathcal{S}_{+}$. 
Thus we define the
nodes of the surfaces $\mathcal{S}_{\pm}$ as
\begin{equation}\label{}
    \mathbf{p}_i^{\pm}=\mathbf{r}_i\pm \Delta \hat{\mathbf{n}}_{i}. \nonumber
\end{equation}
To calculate the exposed area of the surfaces $\mathcal{S}_{\pm}$, 
one places a sphere of radius $R$ tangentially to each triangle on the surfaces
$\mathcal{S}_{\pm}$ at its center of mass, and then to check whether the
sphere overlaps with any of the other points on these two surfaces. If
there is no overlap the triangle is said to be exposed to the solvent.
The following quantity can be calculated for each triangle $(ijk)$ in
the $\mathcal{S}_{\pm}$ surfaces:
\begin{equation}\label{discrBB}
    B^{\pm}_{ijk}=\min_{m\lambda}\vert\mathbf{p}_{ijk}^{\pm}
    {\pm}R\hat{\mathbf{N}}_{ijk}^{\pm}-\mathbf{p}_{m}^{\lambda}\vert, \nonumber
\end{equation}
where 
$\mathbf{p}_{ijk}^{\pm}$ and $\hat{\mathbf{N}}_{ijk}^{\pm}$ is the 
center of mass and the normal of that triangle, respectively, and
$\lambda$ can be $+$ or $-$.
The exposed areas in the surfaces $\mathcal{S}_{\pm}$ are defined as
\begin{equation}\label{discrcontS}
    \Sigma_{\pm}=\frac{1}{2} \sum_{(ijk)}\vert(\mathbf{p}_j^{\pm}-\mathbf{p}_i^{\pm})\times
(\mathbf{p}_k^{\pm}-\mathbf{p}_j^{\pm})\vert\
\Theta\Big[R-B^{\pm}_{ijk}\Big], \nonumber
\end{equation}
where $\Theta()$ is the step function, and the sum is taken over all
triangles on a given surface.

\begin{figure}
\centerline{\includegraphics[width=2.5in]{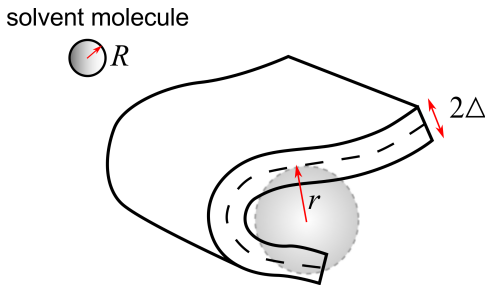}}
\caption{
Cartoon of a thick membrane and a solvent molecule.
Each side of the membrane is at a distance $\Delta$ from the middle surface
(dashed line). Self-avoidance requires that any sphere
tangent to the middle surface at one point and passing through another
point on this surface must have a radius $r \geq \Delta$.
The solvent molecule is considered as a spherical paint brush of radius
$R$. The membrane exposed surface area is defined as the area on both
side of the membrane that is accessible to that spherical paint brush.
}
\label{fig1}
\end{figure}

\section{Model of thick hydrophobic membrane}

Following standard
practice, here we model the thick membrane as a set of points forming a
triangular lattice representing the membrane's middle surface.  The links on
the lattice are fixed and have lengths that are allowed to vary freely between
0.9$l$ and 1.1$l$, where $l$ is the unit length. 
Self-avoidance is implemented through an effective four-body potential (Eq.(9))
to ensure the membrane thickness of $2\Delta$. Due to the effects of
discretization, this procedure is accurate as long as $\Delta/l\geq0.5$.
Note that $\Delta$ provides a length scale that is independent of the
discretization length $l$. The continuum limit can be approached by
decreasing $l$ while keeping $\Delta$ constant.

We consider a membrane that is hexagonal in shape when it is perfectly flat
with the edge length of $L$.
The solvent molecule is modeled as a spherical paint brush with radius $R$.
The interaction between solvent molecules is not considered in our study and
yields a higher order correction. For a hydrophobic solvent, there is a
tendency to minimize the exposed area by appropriate folding of the membrane,
which, at non-zero temperatures, is counteracted by an entropic contribution
promoting relatively flat conformations. An effective self-attraction is thus
introduced through a potential energy: 
\begin{equation}
E = (\varepsilon/l^2)(\Sigma^+ + \Sigma^-) \,,
\end{equation}
where $\varepsilon$ provides an energy unit and $\Sigma^{\pm}$ are the exposed
areas of the $+$ and $-$ sides, respectively.

Monte Carlo simulations are carried out to study the behavior of the membrane
at various temperatures.
In a Monte Carlo move, one node in the triangulated lattice
of the membrane's middle surface is selected and displaced in a random
direction by a random displacement. The magnitude of the displacement is
constrained to be less than 10\% of the lattice constant $l$. The move
is rejected immediately if (1) a new bond length on the lattice due the
displacement is smaller than $0.9l$ or larger than $1.1l$, or (2) the membrane
thickness constraint is violated. 
Otherwise, the move is accepted with probability
$P=\min[1,\exp(-\Delta E/k_B T)]$, where $T$ is temperature, and $\Delta E$ is a
change in the effective energy.
A parallel tempering scheme with 16 to 20 replicas is adopted to get efficient
sampling of the conformational space.

\section{Large vs. small solvent size}

\begin{figure}
\centerline{\includegraphics[width=3.4in]{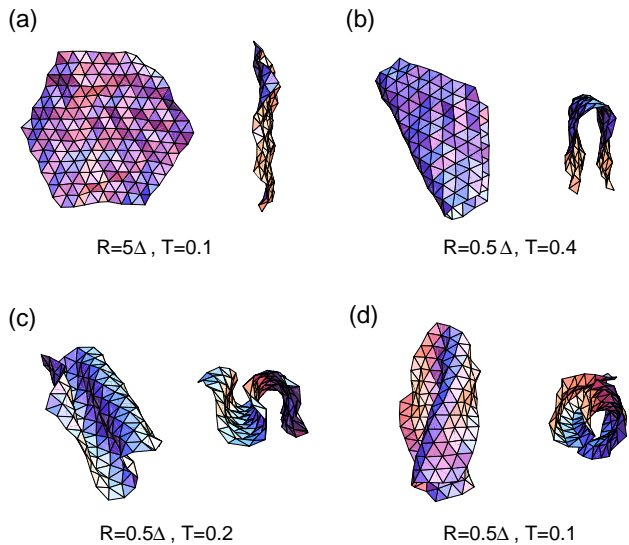}}
\caption{
Conformations of a hydrophobic membrane immersed in water.
The membrane has an edge length of $L=7l$ and the thickness $2\Delta=1.6\,l$.
(a) Two
views of a conformation found at $T=0.1\varepsilon/k_B$ for a membrane
when the solvent radius, $R=5\Delta$, is large. The triangular lattice
shown indicates the membrane's middle layer.  The membrane is fully
buried due to the local crinkling mechanism. Note the the membrane
surface is rough but the overall shape is flat. (b--d) Three typical
folded conformations at low temperatures for a membrane in the small
solvent regime. The solvent radius is $R=\Delta/2$, and the
temperatures are $T=0.4\varepsilon/k_B$ (b), $0.2\varepsilon/k_B$ (c),
and $0.1\varepsilon/k_B$ (d) as indicated. The buried fractions are
0.31, 0.37 and 0.44 for the three cases (b, c, d), respectively.
Note that folding occurs predominantly along one axis.
}
\label{fig2}
\end{figure}

\begin{figure}
\centerline{\includegraphics[width=3.in]{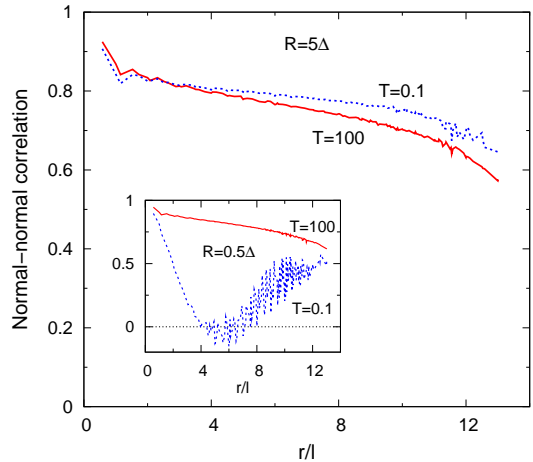}}
\caption{
Correlation between normal vectors.  Dependence of the correlation
between normal vectors, $\langle{\bf n}_x {\bf n}_{x+r}\rangle -
\langle{\bf n}_x\rangle \langle{\bf n}_{x+r}\rangle$, on $r$, the
distance between them, are shown for the large solvent ($R=5\Delta$)
and small solvent ($R=0.5\Delta$) cases. The small solvent case is
shown as inset figure. The membrane has the same characteristics as
the one shown in Fig. \ref{fig2}.  The correlation functions are shown for two
different temperatures: very high, $T=100\varepsilon/k_B$ (solid line) and low,
$T=0.1\varepsilon/k_B$ (dotted line), as indicated.  The data shown for each
temperature are calculated from 100 statistically independent
conformations sampled in parallel tempering Monte Carlo simulations
under equilibrium conditions.
The oscillation in the correlation function for small solvent size at low
temperature reflects the folding along a preferential axis.
Note that in the large solvent case, the correlation, at fixed
distance, is larger as temperature decreases demonstrating a
stiffening of the membrane due to local crinkling.
}
\label{fig3}
\end{figure}

Figs. \ref{fig2} show typical conformations of a membrane in the large and
small solvent size regimes. At high temperatures, the entropy dominates, there
is no tendency
for minimizing the area exposed to the solvent, and the conformations in both
regimes are flat. On lowering the temperature, the membranes tend to become
compact by minimizing their exposed area to the solvent but their behaviors are
remarkably different for the two cases of small and large solvent molecule
sizes.  The correlations between the normal vectors of the membranes also show
a major difference between the low temperature conformations in the large and
small solvent regimes (Fig. \ref{fig3}).

When $R$ is larger than $\Delta$, the low temperature phase is crinkled and
contains many different conformations. The crinkling  is essentially a local
deformation that does not violate the local bending constraint of a thick
membrane yet excludes the large paint brush from accessing the surface.
Locally, below the persistence length, the membrane is {\it essentially flat}
yet crinkled. At the lowest temperature, the crinkled conformations are almost
completely buried (Fig. \ref{fig2}a). We have verified that, for the same
solvent-induced energy function, the impenetrable plaquette model
\cite{Bowick}, behaves similarly to the thick membrane
in the large solvent molecule limit - i.e. the thickness is not relevant when
the solvent molecules are large. Interestingly, for large solvent size, the
local crinkling mechanism induces a stiffening of the membrane in the flat
phase (Fig. \ref{fig3}).

When $R$ is considerably smaller than $\Delta$, the membrane exhibits a
transition from the high temperature flat phase to the low temperature folded
phase characterized by a few different folded geometries (Fig. \ref{fig2}b,c,d). The
membrane undergoes a spontaneous symmetry breaking within the plane of the
membrane. The folding of the membrane occurs multiple times along a dominant
folding axis.

\section{Ribbons in 2-D}

\begin{figure}
\centerline{\includegraphics[width=3.in]{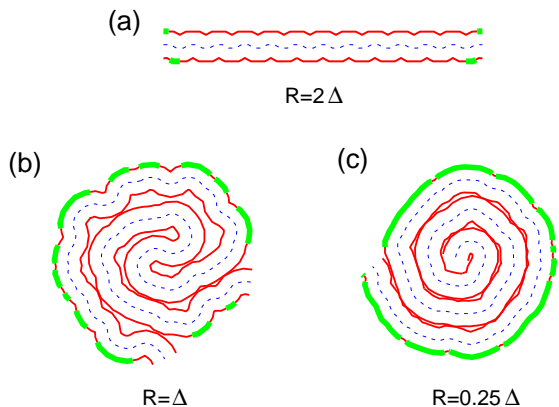}}
\caption{
Conformations of a solvophobic ribbon in 2-D in the presence of a
solvent. In the regime where the membrane is folded (Fig. \ref{fig2}) this is a
simpler (dimensionally reduced) representation. The ribbons shown are
of thickness $2\Delta$ and length $L=25\Delta$ (a) and $L=75\Delta$
(b, c). The solvent is comprised of circular discs of radius $R$ whose
value is indicated. The ribbon axes are shown as dashed lines.
The ribbon perimeters are shown in thinner (thicker)
solid line depending on whether they are buried from (exposed to) the solvent.
The buried fractions are 0.94 (a), 0.82 (b), and 0.72 (c) for the
three cases shown.  Conformations (b) and (c) are the ground states
obtained in Monte Carlo simulations that seek to minimize the exposed
perimeter of the ribbons.  Conformation (c) corresponds to the familiar
rolled carpet conformation of membranes in 3-D in which the normal
vectors lie isotropically within the plane perpendicular to the
folding axis.
}
\label{fig4}
\end{figure}

We have carried out also simulations of solvophobic ribbons in 2-$D$ (Fig.
\ref{fig4}).  The ribbons have a thickness of $2\Delta$ and the solvent
molecule in this case
is considered as circular disc of radius $R$.  We consider a discretized ribbon
axis described by a set of equally spaced points separated by $b=\Delta/2$. The
self-avoidance constraint is imposed by requiring that none of the radii of the
circles passing through each of the triplets of the points on the ribbon axis
is smaller than $\Delta$. All triplets of points are considered.  Like for
membranes, we found two regimes related to the solvent size. For $R > \Delta$,
the ground state of the ribbon is a straight conformation with local crinkling
(Fig. \ref{fig4}a).  For $R \leq \Delta$, one finds folded conformations with some local
crinkling on the outer perimeter (Fig. \ref{fig4} b,c). The folded conformations of the
ribbon correspond to a folded membrane in a dimensionally reduced
representation.  For sufficiently small solvent size, the ground state of the
system is akin to a rolled carpet (Fig. \ref{fig4}c).

Using simple analytic calculations, we find that when
$R>\Delta(2-b^2/\Delta^2)$, a ribbon can adopt a virtually straight
conformation with local crinkling that buries the perimeter completely. Such a
conformation is shown in Fig. \ref{fig4}a for $R=2\Delta$ (the buried fraction does not
reach 1 only because of edge effects). In an optimal local crinkled
conformation, the radius of curvature is equal to $\Delta$ at all points. In
the continuum, one can approximate a crinkled conformation of the ribbon axis
as being formed by joining semicircles of radius $\Delta$, the first convex
upwards, the second concave upwards, and so on. The buried fraction of such a
continuum crinkled ribbon is given by
$f=1-(2/\pi)\arcsin\left(\frac{2}{2+R/\Delta}\right)$, which converges to $1$
as $R/\Delta \rightarrow \infty$.

\begin{figure}
\centerline{\includegraphics[width=2.8in]{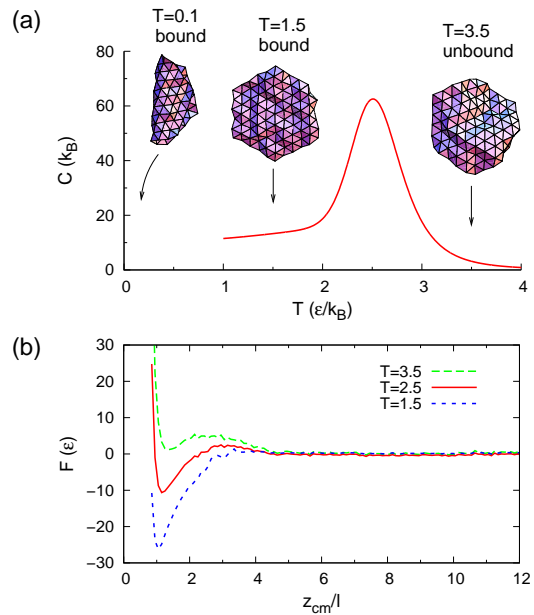}}
\caption{
Adsorption of a membrane on the wall.
(a) Temperature dependence of the specific heat for the system of a
hydrophobic membrane and a wall. The membrane is hexagonal in shape
and has the edge length $L=5l$ and the thickness
$2\Delta=1.2\l$. The solvent radius is $R=\Delta/2$.
Adsorption is induced by the shielding of membrane surface from
the solvent when it is found near the wall.
The adsorption transition is observed at temperature
$T \approx 2.5 \, \varepsilon/k_B$. Membrane conformations are shown
for selected temperatures as indicated. Both the bound and unbound
states at intermediate and high temperatures are flat.
The ground state however is a folded membrane bound the the wall, and
can be found only at very low temperatures ($T=0.1\,\varepsilon/k_B$).
(b) The dependence of the effective free energy, $F$, on the
distance, $z_{cm}$, of the membrane center of mass to the wall for
three temperatures: above, below and at the adsorption transition as
indicated.  The effective free energy is calculated as $F(T,z_{cm}) =
- k_B T \ln P(T,z_{cm})$, where $P(T,z_{cm})$ is the probability of finding the
membrane at the distance $z_{cm}$ of its center of mass to the wall at
temperature $T$.
}
\label{fig5}
\end{figure}

\section{Membrane near a wall}

We have also studied the behavior of a tethered membrane of {\it non-zero
thickness} near a wall which provides a new mechanism for burying the area of a
membrane adsorbed on it. For small solvent molecule size, there is a
competition between the binding to the wall and folding for reducing the
exposed area. Our simulations show that below a temperature, which increases
with membrane size, and is higher than the folding temperature, the membrane is
adsorbed on the wall (Fig. \ref{fig5}). 
An adsorbed membrane would be flat at intermediate 
temperatures and folded at lower temperatures.
Similar considerations suggest that multiple hydrophobic
membranes would adhere to each other, at any temperature, to shield much of the
surface from the solvent yielding an effectively rigid assemblage of membranes.
For large solvent molecules, the adsorption happens below a temperature, that
is higher than the temperature of effective local crinkling. At very low
temperatures, the membrane is crinkled while remaining bound to the wall.

\section{Discussion}

Previous computational studies of membranes have considered, in essence, models
in the infinitesimal thickness limit or 2-d array of spherical particles. The
key finding of studies of traditional models (for a review see
Ref.\cite{Wiese}) is that a self-avoiding tethered membrane always exists
in a flat phase even without any bending rigidity
\cite{Bowick,Abraham89,Grest}. The inclusion of self-attraction
\cite{Kardar1991} leads to several folded phases at intermediate temperatures
along with hints of a crumpled phase \cite{Liu1992} and an isotropic collapsed
phase at low temperatures. A recent study \cite{Starr} with a Lennard-Jones
potential yielded folded sheets and cylindrical conformations.  The crumpled
state has been also studied for an elastic sheet confined in an impenetrable
sphere \cite{Witten} or folded under external forces \cite{Gompper}. Studies of
self-intersecting membranes (or phantom membranes) predict a crumpled phase in
which the membrane's radius of gyration, $R_g$, grows as $\sqrt{\ln L}$, where
$L$ is the membrane's linear size \cite{Gross,Duplantier}. With an increase in
bending rigidity, phantom membranes undergo a second-order transition from the
crumpled phase to the flat phase \cite{Kantor-Nelson,Lubensky}.  The flat,
crumpled, and collapsed phases have been observed experimentally in red cell
membrane skeletons \cite{Schmidt1993}, graphite oxide membrane
\cite{Tanaka1992}, and molybdenum disulfide \cite{Chianelli1979}.

Here we show that incorporating non-zero thickness yields qualitatively new
behavior of a hydrophobic membrane seeking to minimize the area exposed to the
surrounding solvent molecules.  Our simulations show that in the small solvent
molecule regime, the thickness of the membrane introduces a spontaneous
symmetry breaking within the plane of the membrane.  The folding of the
membrane occurs multiple times along a dominant folding axis. This is akin to
the everyday experience of the greater ease of folding a sheet of paper in a
fluted manner compared to first folding the paper along one axis followed by
folding the now two layered sheet along a perpendicular axis and so on. A rolled carpet
conformation is also effective in burying much of the surface from the solvent
molecules (Fig. \ref{fig4}c). Note that the folded structures obtained here are
similar to the those in the tubule phase \cite{Radzihovsky} -- even in the
absence of any anisotropic bending energy, we observe spontaneous symmetry
breaking. For a large membrane, one obtains uncoordinated folding at the edges
into metastable conformations.
The structures of the folded thick membrane are qualitatively similar to those
observed in molybdenum disulfide films \cite{Chianelli1979} and in folding of
viscous sheets and filaments \cite{Mahadevan}.

For large solvent molecules, the local crinkling mechanism effectively shields
the surfaces of the membrane from the solvent, while preserving the overall
flat topology. Similar local deformations are observed in real membranes.
Wrinkled rigid structures have been observed experimentally upon cooling
partially polymerized phospholipid membranes \cite{Bensimon1991} and have been
interpreted as resulting from quenched curvature disorder leading to a glassy
phase. Studies of wrinkled patterns of polymer films \cite{Menon} have
indicated that thickness plays a role in the formation of these patterns.

We note that the our study considers paint brush sizes in two distinct regimes:
smaller than or larger than the membrane thickness. Membranes thicker than the
size of a water molecule ($\approx 2.8\AA$) are quite common.
The other limit is
obtained when the role of solvent is played by large complexes such as proteins
(3-10 nm), t-RNA (7 nm), antibodies (12 nm) and ribosomes (20-30 nm).
Interestingly, for a relatively dilute concentration of these complexes,
even of the order of 10\% in volume fraction according to Asakura-Oosawa theory
\cite{Oosawa}, the entropy depletion effect alone is sufficient to create a
tendency to fold the membrane leading to minimization of the exposed area.

\acknowledgments
We are indebted to Mehran Kardar, Tom Lubensky, and Michael Plischke for 
useful comments on the manuscript. This article was supported by the Vietnam
National Foundation for Science and Technology Development (NAFOSTED)
grant 103.01-2010.11 and Prin 2007.

\end{document}